# Effects of annealing time on structural and magnetic properties of $L1_0$-FePt nanoparticles synthesized by the SiO$_2$-nanoreactor method

Yoshinori Tamada, Yasumasa Morimoto, Shinpei Yamamoto, Mikio Takano, Saburou Nasu and Teruo Ono

*Institute for Chemical Research, Kyoto University, Uji 611-0011, Japan*

**Abstract**

We investigated effects of annealing time on structural and magnetic properties of the $L1_0$-FePt nanoparticles synthesized by the "SiO$_2$-nanoreacter" method. The magnetization and powder X-ray diffraction studies revealed that the annealing at 900 °C for 9 hr could convert all of the fcc-nanoparticles to the well-crystallized $L1_0$ structure with a large coercivity while keeping their particle size. Such monodisperse and highly crystalline $L1_0$-FePt nanoparticles are a promising material for the realization of ultra-high density recording.

**Introduction**

FePt nanoparticles synthesized by chemical solution-based methods attract much attention as one of the most promising candidates for the future recording media with ultra-high densities beyond 1 Tbit/inch$^2$ [1,2]. However, the chemical methods can produce only disordered face-centered cubic (fcc) or partially ordered $L1_0$-FePt nanoparticles [3,4], and thus a post thermal annealing is necessary to transform them into the desired, well-ordered $L1_0$ structure. This post annealing results in coalescence and coarsening of the nanoparticles, leading to difficulties in fabricating desirable arrays on a substrate.

Recently we have succeeded in solving these problems by developing a new synthetic strategy named "SiO$_2$-nanoreactor" method [5,6]. Previous studies revealed that annealing at 900 °C for 1 hr could convert almost all of the fcc-FePt nanoparticles to the well-crystallized $L1_0$ structure with a high coercivity although only minor part of the sample remained unconverted.

Here, we report effects of annealing time on structural and magnetic properties of the nanoparticles synthesized by the SiO$_2$-nanoreactor method by means of SQUID magnetometry and powder X-ray diffractometry (XRD).

**Experimental**

$L1_0$-FePt nanoparticles were prepared according to the reported method [5,6]. The SiO$_2$-coated fcc-FePt nanoparticles (6.5 nm in diameter) were annealed at 900 °C for various periods of time in flowing H$_2$(5%)/Ar(95%) gas to convert them to the $L1_0$ structure. Elemental composition of the FePt nanoparticles was determined to be Fe$_{48}$Pt$_{52}$ by using an atomic absorption spectrometer (Shimadzu, AA-6300). The XRD profile collected by using Cu $K_\alpha$ radiation ($\lambda$ = 0.154 nm) (Rigaku, RINT2500).

Magnetic properties were characterized by using a SQUID magnetometer (Quantum Design, MPMS XL).

**Results and Discussion**

Figure 1 shows the room temperature hysteresis loops for the $SiO_2$-coated nanoparticles annealed at 900 ℃ for various periods of time. Here, $M_s$ represents the overall sample magnetization at 50 kOe. Coercivity of the FePt nanoparticles annealed for 1hr reaches as large as 13 kOe, confirming Fe and Pt atoms forms well-ordered $L1_0$ structure. However, the hysteresis loop shows a kink at low magnetic field. This indicates the sample includes magnetically soft phase, *i.e.*, the fcc phase, in addition to the $L1_0$ phase. The amount of the fcc phase was estimated to be 15 % of all the nanoparticles from the decrease of magnetization at the kink. It is notable that the decrease of magnetization at the kink became smaller (7.5 % for 3hr, 0 % for 9hr) and the coercivity became larger as the annealing time increased. These features indicate that the fcc-nanoparticles remained after annealing for 1 hr could be converted to the $L1_0$-phase upon further annealing (Table 1).

Figure 2 shows a series of XRD patterns from $2\theta = 38$ º to 44 º of the $SiO_2$-coated FePt nanoparticles annealed 900 ℃ at for various periods of time. Though there are both the $L1_0$- and the fcc-phases (111) main reflection peaks in this range, it is difficult to directly estimate relative amount of them because of the broadening of peak width arising from the small size of the nanoparticles. Therefore we divided the observed signal into two peaks of the $L1_0$- and the fcc- phases by assuming that they have Lorentzian form. A good fit to the signal could be obtained by assuming the constant size of the $L1_0$- and fcc-nanoparticles being 7.1 and 4.5 nm, respectively. We can estimate the relative amount of the $L1_0$- and fcc-nanoparticles from these parameters, which are shown in Table 1. The XRD studies also revealed the decrease of the amount of the fcc-nanoparticles with increasing the annealing time. It is worth noting that the results from the XRD agree well with the estimation from the magnetization measurements.

**Conclusions**

We investigated effects of annealing time on structural and magnetic properties of the nanoparticles synthesized by the $SiO_2$-nanoreactor method. The magnetization and XRD studies revealed that the annealing at 900 $^o$C for 9 hr could convert all of the fcc-nanoparticles to the well-crystallized $L1_0$ structure with a large coercivity while keeping their particle size. We expect that such monodisperse and highly crystalline $L1_0$-FePt nanoparticles are a promising material for the realization of ultra-high density recording.

**Figure Captions**

Figure 1
Room temperature hysteresis loops of the $SiO_2$-coated FePt nanoparticles annealed at 900 ºC for various periods of time.

Figure 2
XRD patterns of the $SiO_2$-coated FePt nanoparticles annealed at 900 ºC for various periods of time.

Table 1

Relative amount of the $L1_0$- and fcc-nanoparticles estimated by XRD patterns of the SiO$_2$-coated FePt nanoparticles annealed for various periods of time.

| annealing time | $L1_0$ (%) | fcc (%) |
|---|---|---|
| 1h | 83 | 17 |
| 3h | 93.4 | 6.6 |
| 9h | 100 | 0 |

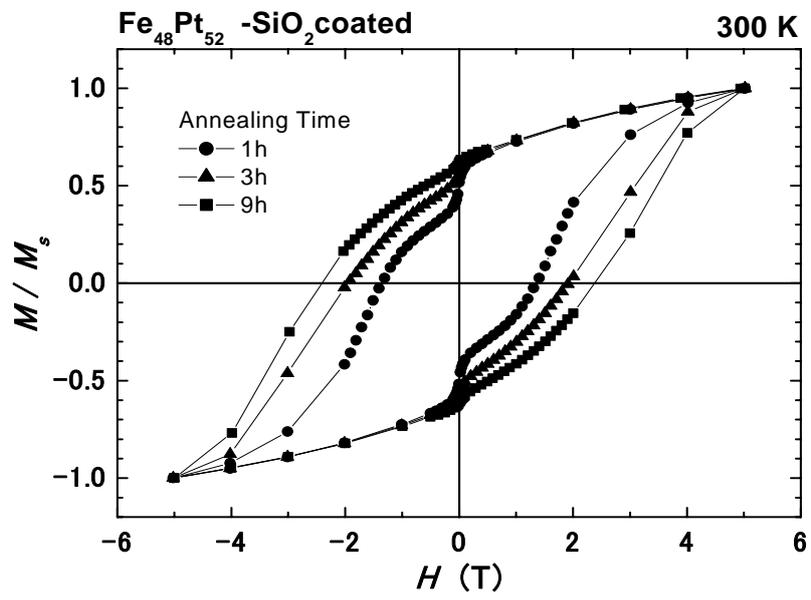

Figure 1
Y. Tamada

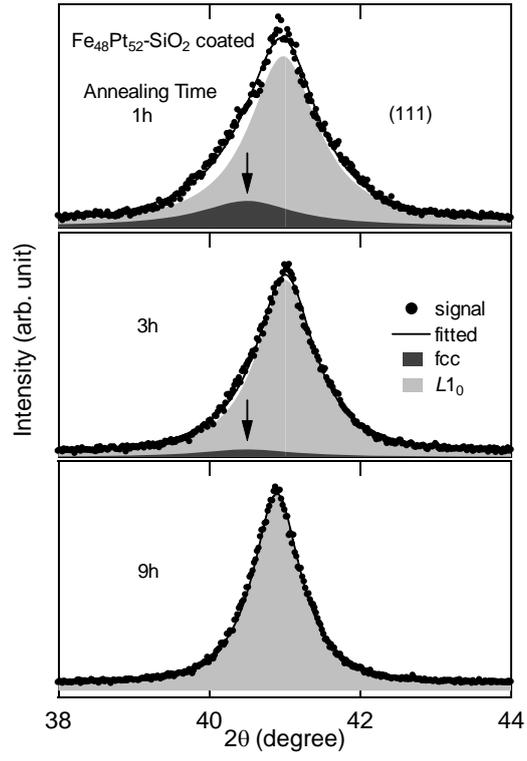

Figure 2
Y. Tamada